\documentclass[preprintnumbers,unsortedaddress,superscriptaddress,notitlepage]{revtex4}
\usepackage{amsfonts}
\usepackage{amsmath}
\usepackage{hyperref}
\usepackage{amssymb}
\usepackage[english]{babel}
\usepackage{graphicx}
\usepackage{epsfig}
\usepackage{bm}
\usepackage{longtable}
\usepackage{verbatim}
\usepackage{longtable}

\newcommand{\mathsym}[1]{{}}

\def\func#1{\mathop{\rm #1}\nolimits}%
\begin{document}

\title{Lepton masses and mixing in $SU\left( 3\right) _{C}\otimes
SU\left(3\right) _{L}\otimes U\left( 1\right) _{X}$ models with a $S_3$
flavor symmetry}
\author{A. E. C\'{a}rcamo Hern\'{a}ndez}
\email{antonio.carcamo@usm.cl}
\affiliation{Universidad T\'{e}cnica Federico Santa Mar\'{\i}a and Centro Cient\'{\i}%
fico-Tecnol\'{o}gico de Valpara\'{\i}so. Casilla 110-V, Valpara\'{\i}so,
Chile. }
\author{E. Catano Mur}
\email{ecatanom@iastate.edu}
\affiliation{Department of Physics and Astronomy, Iowa State University. Ames, Iowa, USA.}
\author{R. Martinez}
\email{remartinezm@unal.edu.co}
\affiliation{Departamento de F\'{\i}sica, Universidad Nacional de Colombia, Ciudad
Universitaria, Bogot\'{a} D.C., Colombia.}

\begin{abstract}
We propose a model based on the gauge group $SU(3)_{C}\otimes
SU(3)_{L}\otimes U(1)_{X}$ with an extra $S_{3}$ flavor symmetry, which
accounts for the lepton masses and mixing. The small active neutrino masses
are generated via a double seesaw mechanism. In this scenario, the spectrum
of neutrinos presents very light, light and very heavy masses. The model
predicts a quasidegenerate normal hierarchy active neutrino mass spectrum
and the relation $\Delta m_{21}^{2}\ll \Delta m_{31}^{2}$ arises from
effective six-dimensional operators. The obtained neutrino mixing parameters
are in agreement with the neutrino oscillation experimental data. We find CP
violation in neutrino oscillations with a Jarlskog invariant of about $%
10^{-2}$. 

\end{abstract}

\maketitle

%


\section{Introduction}

The experimental confirmation of the electroweak symmetry breaking (EWSB)
sector of the Standard Model (SM) given by the discovery of the Higgs Boson
at the LHC \cite{atlashiggs,cmshiggs} has concreted its great success in
describing electroweak phenomena. However, the SM does not explain neither
the pattern of fermion masses and mixing nor the existence of three
generations of fermions. In consequence, to address these issues it is
necessary to consider a more fundamental theory. The existing pattern of
fermion masses goes over a range of five orders of magnitude in the quark
sector and a much wider range when neutrinos are included. While in the
quark sector the mixing angles are small, in the lepton sector two of the
mixing angles are large, and one is small; this suggests that the
corresponding mechanisms for masses and mixings should be different.
Experiments with solar, atmospheric and reactor neutrinos \cite%
{PDG,An:2012eh,Abe:2011sj,Adamson:2011qu,Abe:2011fz,Ahn:2012nd} have brought
evidence of neutrino oscillations caused by nonzero mass. 
The global fits of the available data from the Daya Bay \cite{An:2012eh},
T2K \cite{Abe:2011sj}, MINOS \cite{Adamson:2011qu}, Double CHOOZ \cite%
{Abe:2011fz} and RENO \cite{Ahn:2012nd} neutrino oscillation experiments,
constrain the neutrino mass squared splittings and mixing parameters \cite%
{Tortola:2012te}.

Models with an extended gauge symmetry are frequently used to tackle the
limitations of the SM. In particular, those based on the $SU(3)_{C}\otimes
SU(3)_{L}\otimes U(1)_{X}$ gauge symmetry, called 331 for short, can explain
the origin of fermion generations thanks to the introduction of a family
nonuniversal $U(1)_{X}$ symmetry \cite{331-pisano,M-O}. Specific
realizations of 331 models have several appealing features. %
%
First, the three family structure in the fermion sector is a consequence of
the chiral anomaly cancellation \cite{anomalias} and the asymptotic freedom
in QCD. Second, the large mass splitting between the heaviest quark family
and the two lighter ones can be explained since the former is in a different 
$U(1)_{X}$ representation \cite{third-family}. Third, these models include a
natural Peccei-Quinn symmetry, that sheds light on the strong-CP problem 
\cite{PC}. Finally, versions with heavy sterile neutrinos have cold dark
matter candidates as weakly interacting massive particles (WIMPs) \cite%
{DM331}. We consider 331 models with a scalar sector composed of three $%
SU(3)_{L}$ scalar triplets, where one heavy triplet field acquires a vacuum
expectation value (VEV) at a high energy scale, $v_{\chi }$, responsible for
breaking the symmetry $SU(3)_{L}\otimes U(1)_{X}$ down to the SM electroweak
gauge group $SU(2)_{L}\otimes U(1)_{Y}$; and two lighter triplets get VEVs $%
v_{\rho }$ and $v_{\eta }$ at the electroweak scale, thus triggering the
EWSB. 

On the other hand, discrete flavor symmetries are important ingredients in
models of particle masses and mixing, and many of them have been considered
to resolve the fermion mass hierarchy; for recent reviews see Refs. \cite{FlavorSymmRev,King:2013eh,Altarelli:2010gt,Ishimori:2010au}. In particular
the $S_{3}$ flavor symmetry is a very good candidate for explaining the
prevailing pattern of fermion masses and mixing. The $S_{3}$ discrete
symmetry is the smallest non-Abelian discrete symmetry group having three
irreducible representations (irreps), explicitly two singlets and one
doublet irreps \cite{s3pheno}. 
Since two of the three $SU(3)_{L}$ scalar triplets of the 331 models belong
to the same $U(1)_{X}$ representation while the third is in a different one,
the scalar fields can be arranged into doublet and non trivial $S_{3}$
singlet irreps. Regarding charged leptons, we accommodate left- and right-
handed leptons as well as one heavy Majorana neutrino into $S_{3}$ singlet
representations, and the remaining two heavy Majorana neutrinos into a $%
S_{3} $ doublet representation. We assume that the heavy Majorana neutrinos
have masses much larger than the TeV scale, so that the hierarchy $M_{R}\gg
v_{\chi }\gg v_{\rho },v_{\eta }$ is fulfilled, implying that the small
active neutrino masses are generated via a double seesaw mechanism. This
mechanism does not include any exotic charges, neither in the fermionic nor
in the scalar sector \cite{catano}. We predict a quasidegenerate normal
hierarchy active neutrino mass spectrum and the relation $\Delta
m_{21}^{2}\ll \Delta m_{31}^{2}$ results from effective six-dimensional
Yukawa terms. 

This paper is organized as follows. In Sec. \ref{331model} we explain
some theoretical aspects of the 331 model with $\beta =-\frac{1}{\sqrt{3}}$
and its particle content, as well as the particle assignments under doublet
and singlet $S_{3}$ representations, in particular in the fermionic and
scalar sector. 
In Sec. \ref{lepton} we focus on the discussion of neutrino masses and
mixing and give our corresponding results. Conclusions are given Sec. \ref%
{conclusions}. In the appendices we present several technical details:
Appendix \ref{appA} gives a brief description of the $S_{3}$ group; Appendix %
\ref{appC} shows the diagonalization of the neutrino mass matrix.


\section{A $SU\left( 3\right) _{C}\otimes SU\left( 3\right) _{L}\otimes
U\left( 1\right) _{X}\otimes S_{3}$ model with $\protect\beta =\frac{-1}{%
\protect\sqrt{3}}$}
\label{331model} 

\subsection{Particle content}


We consider a 331 model with $\beta =-\frac{1}{\sqrt{3}}$ \cite%
{catano,331-2hdm,Hernandez:2013hea}, where the electric charge is defined in
terms of $SU(3)$ generators and the identity by
\begin{equation}
Q=T_{3}-\frac{1}{\sqrt{3}}T_{8}+XI,
\end{equation}%
with $I=Diag(1,1,1)$, $T_{3}=\frac{1}{2}Diag(1,-1,0)$ and $T_{8}=(\frac{1}{2\sqrt{3}}%
)Diag(1,1,-2)$. To avoid chiral anomalies, fermions are assigned to the
following $(SU(3)_{C},SU(3)_{L},U(1)_{X})$ left- and right-handed
representations: 
\begin{align}
Q_{L}^{1,2}& =%
\begin{pmatrix}
D^{1,2} \\ 
-U^{1,2} \\ 
J^{1,2} \\ 
\end{pmatrix}%
_{L}:(3,3^{\ast },0), & & 
\begin{cases}
D_{R}^{1,2}:(3^{\ast },1,-1/3) \\ 
U_{R}^{1,2}:(3^{\ast },1,2/3) \\ 
J_{R}^{1,2}:(3^{\ast },1,-1/3) \\ 
\end{cases}%
,  \notag \\
Q_{L}^{3}& =%
\begin{pmatrix}
U^{3} \\ 
D^{3} \\ 
T \\ 
\end{pmatrix}%
_{L}:(3,3,1/3), & & 
\begin{cases}
U_{R}^{3}:(3^{\ast },1,2/3) \\ 
D_{R}^{3}:(3^{\ast },1,-1/3) \\ 
T_{R}:(3^{\ast },1,2/3) \\ 
\end{cases}%
,  \notag \\
L_{L}^{1,2,3}& =%
\begin{pmatrix}
\nu ^{1,2,3} \\ 
e^{1,2,3} \\ 
\left( \nu ^{1,2,3}\right) ^{c} \\ 
\end{pmatrix}%
_{L}:(1,3,-1/3), & & 
\begin{cases}
e_{R}^{1,2,3}:(1,1,-1) \\ 
N_{R}^{1,2,3}:(1,1,0) \\ 
\end{cases}%
,
\end{align}%
where $U_{L}^{i}$ and $D_{L}^{i}$ for $i=1,2,3$ are three up- and down-type
quark components in the flavor basis, while $\nu _{L}^{i}$ and $e_{L}^{i}$
are the neutral and charged leptons. The right-handed components transform
as singlets under $SU(3)_{L}$ with $U(1)_{X}$ quantum numbers corresponding
to the electric charges.

Additionally, the model includes heavy fermions with the following
properties: a single flavor quark $T$ with electric charge $2/3$, two flavor
quarks $J^{2,3}$ with charge $-1/3$, three neutral Majorana leptons $\left(
\nu ^{1,2,3}\right) _{L}^{c}$ and three right-handed Majorana leptons $%
N_{R}^{1,2,3}$. 
\newline
The scalar sector consists of a triplet field $\chi $, which provides the
masses to the new heavy fermions, and two triplets $\rho $ and $\eta $,
which give masses to the SM fermions at the electroweak scale. The $\left(
SU(3)_{L},U(1)_{X}\right) $ group structure of the scalar fields is: 
\begin{eqnarray}
\chi &=&%
\begin{pmatrix}
\chi _{1}^{0}+\frac{1}{\sqrt{2}}w_{\chi }e^{i\varphi _{\chi }} \\ 
\chi _{2}^{-} \\ 
\frac{1}{\sqrt{2}}(\upsilon _{\chi }+\xi _{\chi }\pm i\zeta _{\chi }) \\ 
\end{pmatrix}%
:(3,-1/3)  \notag \\
\rho &=&%
\begin{pmatrix}
\rho _{1}^{+} \\ 
\frac{1}{\sqrt{2}}(\upsilon _{\rho }+\xi _{\rho }\pm i\zeta _{\rho }) \\ 
\rho _{3}^{+} \\ 
\end{pmatrix}%
:(3,2/3)  \notag \\
\eta &=&%
\begin{pmatrix}
\frac{1}{\sqrt{2}}(\upsilon _{\eta }+\xi _{\eta }\pm i\zeta _{\eta }) \\ 
\eta _{2}^{-} \\ 
\eta _{3}^{0}+\frac{1}{\sqrt{2}}w_{\eta }e^{i\varphi _{\eta }}%
\end{pmatrix}%
:(3,-1/3).  \label{fermion_spectrum}
\end{eqnarray}

The electroweak symmetry breaking (EWSB) mechanism follows 
\begin{equation*}
{SU(3)_{L}\otimes U(1)_{X}\xrightarrow{\langle \chi \rangle}}{%
SU(2)_{L}\otimes U(1)_{Y}}{\xrightarrow{\langle \eta \rangle,\langle \rho
\rangle}}{U(1)_{Q}},
\end{equation*}%
where the vacuum expectation values satisfy the hierarchy $v_{\chi }\gg
v_{\eta },v_{\rho }\gg w_{\chi }$, $w_{\eta }.$ Notice that we have
introduced nonvanishing complex vacuum expectation values in the first and
third components of the $\chi $ and $\eta $ triplets, respectively, as done
in Refs. \cite{Dong:2008sw,Dong:2010gk}.

In order to reduce the number of parameters in the Yukawa and scalar sectors
of the $331$ Lagrangian, we impose a $S_{3}$ flavor symmetry for fermions
and scalars, making $SU(3)_{C}\otimes SU\left( 3\right) _{L}\otimes U\left(
1\right) _{X}\otimes S_{3}$ the full symmetry of our model. Apart from
easily accommodating maximal mixing through its doublet representation, the $%
S_{3}$ discrete group has two different singlet representations crucial for
reproducing the fermion masses \cite{s3pheno}. 
The scalar fields are assigned into doublet and singlet representations of $%
S_{3}$ as follows,
\begin{equation}
\Phi =\left( \eta ,\chi \right) \sim \mathbf{2},\qquad \rho \sim \mathbf{1}%
^{\prime },  \label{1}
\end{equation}%
whereas the leptons transform under $S_{3}$ as 
\begin{equation}
L_{L}^{1,2,3}\sim \mathbf{1},\qquad e_{R}^{1,2,3}\sim \mathbf{1}^{\prime
},\qquad N_{R}^{1}\sim \mathbf{1},\qquad N_{R}=\left(
N_{R}^{2},N_{R}^{3}\right) \sim \mathbf{2}.  \label{leptonfields}
\end{equation}%
The corresponding $S_{3}$ assignments for quarks as well as the quark masses
and mixing are studied in detail in the $SU(3)_{C}\otimes SU\left( 3\right)
_{L}\otimes U\left( 1\right) _{X}\otimes S_{3}$ model of Ref. \cite%
{Hernandez:2013hea}.

With the above spectrum, we obtain the following Yukawa terms for the lepton
sector invariant under $S_{3}$: 
\begin{eqnarray}
-\mathcal{L}_{Y}^{\left( L\right) } &=&h_{\rho e}^{\left( L\right) }%
\overline{L}_{L}\rho e_{R}+h_{\Phi }^{\left( L\right) }\overline{L}%
_{L}\left( \Phi N_{R}\right) _{\mathbf{1}}+\frac{1}{2}m_{N}^{\left( 1\right)
}\overline{N}_{R}^{1}N_{R}^{1C}+\frac{1}{2}m_{N}\left( \overline{N}%
_{R}N_{R}^{C}\right) _{\mathbf{1}}+\frac{h_{\Phi }^{\left( N\right) }}{%
\Lambda }\overline{N}_{R}^{1}N_{R}^{C}\left( \Phi \Phi ^{\dagger }\right) _{%
\mathbf{2}}  \notag \\
&&+\frac{h_{\rho }}{\Lambda ^{2}}\overline{L}_{L}^{a}\left(
L_{L}^{C}\right) ^{b}\rho ^{c}\varepsilon _{abc}\left( \Phi ^{\dagger }\Phi
\right) _{\mathbf{1}^{\prime }}+\text{H.c.}  \label{lyl}
\end{eqnarray}%
%
%
%
%
%
Note that the heavy Majorana neutrinos $N_{R}^{2}$ and $N_{R}^{3}$ belonging
to the same $S_{3}$ doublet have the same mass $m_{N}$, which is in general
different than the mass $m_{N}^{\left( 1\right) }$ of the heavy Majorana
neutrino $\overline{N}_{R}^{1}$. Therefore, the $S_{3}$ flavor symmetry
leads to a heavy Majorana neutrino mass splitting, so that $m_{N}^{\left(
1\right) }=\kappa m_{N}$ where the dimensionless parameter $\kappa $ may
differ from $1$.

In order to see if there are operators of dimension larger than four that
contribute to the neutrino masses, first we consider the bilinear combinations of
two leptonic fields of Eq. (\ref{leptonfields}). In the Yukawa terms given
by Eq. (\ref{lyl}), we have already found the combinations $\overline{L}_{L}N_{R}$, $\overline{N}_{R}^{1}N_{R}^{1C}$, $%
\overline{N}_{R}N_{R}^{C}$, $\overline{L}_{L}L_{L}^{C}$ and $\overline{N}%
_{R}^{1}N_{R}^{C}$; the only  combination missing is
 $\overline{L}_{L}N_{R}^{1}$.  
Table \ref{bilinearcombinations} shows the $%
SU\left( 3\right) _{L}\otimes S_{3}$ invariant operators with dimension
larger than four built from these bilinears that
could contribute to the neutrino masses. Only the
terms $\frac{1}{\Lambda }\overline{N}_{R}^{1}N_{R}^{C}\left( \Phi \Phi
^{\dagger }\right) $ and $\frac{1}{\Lambda ^{2}}\overline{L}%
_{L}L_{L}^{C}\rho \left( \Phi ^{\dagger }\Phi \right) $ have vanishing $%
U(1)_{X}$ charge and thus they are invariant under the group $SU\left(
3\right) _{L}\otimes U\left( 1\right) _{X}\otimes S_{3}$. The five-dimensional Yukawa term gives a subleading contribution to the heavy
Majorana neutrino masses. That contribution is supressed by factor of about $\sim\frac{\upsilon _{\chi }^{2}}{\Lambda }<\!\!<m_{N}$ where $\Lambda $ is the
cutoff of our model.%

\begin{table}[tbh]
\begin{tabular}{|c|c|c|c|}
\hline
Operator & $SU(3)_L$ & $S_3$ & $U(1)_X$ \\ \hline
&  &  &  \\ 
\hspace{0.3cm}$\frac1{\Lambda}\overline{L}_{L}N^1_{R}\left(\Phi^{\dagger}%
\Phi^{\dagger}\right)$\hspace{0.3cm} & \hspace{0.3cm}Invariant\hspace{0.3cm}
& \hspace{0.3cm}Invariant\hspace{0.3cm} & \hspace{0.3cm} $\ne 0$\hspace{0.3cm%
} \\ 
&  &  &  \\ \hline
&  &  &  \\ 
$\frac1{\Lambda}\overline{N}^1_{R}N^C_R\left(\Phi\rho^{\dagger}\right)$ & 
Invariant & Invariant & $\ne 0$ \\ 
&  &  &  \\ \hline
&  &  &  \\ 
$\frac1{\Lambda}\overline{N}^1_{R}N^C_R\left(\Phi\Phi^{\dagger}\right)$ & 
Invariant & Invariant & $=0$ \\ 
&  &  &  \\ \hline
&  &  &  \\ 
$\frac1{\Lambda^2}\overline{L}_{L}N^1_{R}\rho\left(\Phi^{\dagger}\Phi\right)$
& Invariant & Invariant & $\ne 0$ \\ 
&  &  &  \\ \hline
&  &  &  \\ 
$\frac1{\Lambda^2}\overline{L}_{L}L^C_L\rho\left(\Phi^{\dagger}\Phi\right)$
& Invariant & Invariant & $=0$ \\ 
&  &  &  \\ \hline
\end{tabular}%
\caption{$SU\left( 3\right) _{L}\otimes S_{3}$ invariant operators with
dimension larger than four that could contribute to the neutrino masses.}
\label{bilinearcombinations}
\end{table}

\subsection{Scalar potential}

The scalar potential of the model is constructed with the $S_{3}$ doublet $%
\Phi =\left( \eta ,\chi \right) $ and the nontrivial $S_{3}$ singlet $\rho $
fields, in the way invariant under the group $SU(3)_{C}\otimes SU\left(
3\right) _{L}\otimes U\left( 1\right) _{X}\otimes S_{3}$. It is given by:%
\begin{align}
V_{H}& =\mu _{\rho }^{2}(\rho ^{\dagger }\rho )+\mu _{\Phi }^{2}\left( \Phi
^{\dagger }\Phi \right) _{\mathbf{1}}+\lambda _{1}(\rho ^{\dagger }\rho
)(\rho ^{\dagger }\rho )+\lambda _{2}\left( \Phi ^{\dagger }\Phi \right) _{%
\mathbf{1}}\left( \Phi ^{\dagger }\Phi \right) _{\mathbf{1}}+\lambda
_{3}\left( \Phi ^{\dagger }\Phi \right) _{\mathbf{1}^{\prime }}\left( \Phi
^{\dagger }\Phi \right) _{\mathbf{1}^{\prime }}  \notag \\
& +\lambda _{4}\left( \Phi ^{\dagger }\Phi \right) _{\mathbf{2}}\left( \Phi
^{\dagger }\Phi \right) _{\mathbf{2}}+\lambda _{5}(\rho ^{\dagger }\rho
)\left( \Phi ^{\dagger }\Phi \right) _{\mathbf{1}}+\lambda _{6}\left( (\rho
^{\dagger }\Phi )\left( \Phi ^{\dagger }\rho \right) \right) _{\mathbf{1}} 
\notag \\
& +f\left[ \varepsilon ^{ijk}\left( \Phi _{i}\Phi _{j}\right) _{\mathbf{1}%
^{\prime }}\rho _{k}+\text{H.c.}\right],  \label{5}
\end{align}

where $\Phi _{i}=\left( \eta _{i},\chi _{i}\right) $ is a $S_{3}$ doublet
with $i=1,2,3$ and all parameters of the scalar potential have to be real.

We softly break the $S_{3}$ symmetry in the quadratic term of the scalar
potential since the vacuum expectation values of the scalar fields $\eta $
and $\chi $ contained in the $S_{3}$ doublet $\Phi $ satisfy the hierarchy $%
v_{\chi }\gg v_{\eta }$. %
Then, considering the quadratic $S_{3}$ soft-breaking terms $\left( \mu
_{\eta }^{2}-\mu _{\chi }^{2}\right) \left( \eta ^{\dagger }\eta \right) $
and $\mu _{\eta \chi }^{2}\left( \chi ^{\dagger }\eta \right) +\text{H.c.}$ using
the multiplication rules of the $S_{3}$ group, the scalar potential can be
written in terms of the three scalar triplets as follows: 
\begin{align}
V_{H}& =\mu _{\rho }^{2}(\rho ^{\dagger }\rho )+\mu _{\eta }^{2}\left( \eta
^{\dagger }\eta \right) +\mu _{\chi }^{2}\left( \chi ^{\dagger }\chi \right)
+\mu _{\eta \chi }^{2}\left[ \left( \chi ^{\dagger }\eta \right) +\left(
\eta ^{\dagger }\chi \right) \right] +\lambda _{1}(\rho ^{\dagger }\rho
)^{2}+\left( \lambda _{2}+\lambda _{4}\right) \left[ \left( \chi ^{\dagger
}\chi \right) ^{2}+(\eta ^{\dagger }\eta )^{2}\right]  \notag \\
& +\lambda _{5}\left[ (\rho ^{\dagger }\rho )(\chi ^{\dagger }\chi )+(\rho
^{\dagger }\rho )(\eta ^{\dagger }\eta )\right] +2\left( \lambda
_{2}-\lambda _{4}\right) \left( \chi ^{\dagger }\chi \right) \left( \eta
^{\dagger }\eta \right) +2\left( \lambda _{4}-\lambda _{3}\right) \left(
\chi ^{\dagger }\eta \right) \left( \eta ^{\dagger }\chi \right)  \notag \\
& +\lambda _{6}\left[ \left( \chi ^{\dagger }\rho \right) (\rho ^{\dagger
}\chi )+\left( \eta ^{\dagger }\rho \right) (\rho ^{\dagger }\eta )\right]
+\left( \lambda _{3}+\lambda _{4}\right) \left[ \left( \chi ^{\dagger }\eta
\right) ^{2}+\left( \eta ^{\dagger }\chi \right) ^{2}\right]  \notag \\
& +2f\left( \varepsilon ^{ijk}\eta _{i}\chi _{j}\rho _{k}+\text{H.c.}\right) .
\label{V1}
\end{align}

It is worth mentioning that the $S_{3}$ soft-breaking term $\mu _{\eta \chi
}^{2}\left( \chi ^{\dagger }\eta \right) +\text{H.c.}$ is not relevant for the
minimization conditions of the scalar potential as well as for the masses of
the physical scalars.

Considering $f,v_{\chi }\gg v_{\eta },v_{\rho }$, we found in detail in Ref. 
\cite{Hernandez:2013hea} the physical scalar mass eigenstates. The CP-even
scalar mass eigenstates are
\begin{equation}
\left( 
\begin{array}{c}
H_{1}^{0} \\ 
h^{0}%
\end{array}%
\right) \simeq \left( 
\begin{array}{cc}
\cos \alpha & -\sin \alpha \\ 
\sin \alpha & \cos \alpha%
\end{array}%
\right) \left( 
\begin{array}{c}
\xi _{\rho } \\ 
\xi _{\eta }%
\end{array}%
\right) ,\hspace{1cm}\hspace{1cm}H_{3}^{0}\simeq \xi _{\chi }.
\end{equation}

The CP-odd scalar mass eigenstates are
\begin{equation}
\left( 
\begin{array}{c}
A^{0} \\ 
G_{1}^{0} \\ 
G_{3}^{0}%
\end{array}%
\right) =\left( 
\begin{array}{ccc}
\cos \beta & \sin \beta & 0 \\ 
\sin \beta & -\cos \beta & 0 \\ 
0 & 0 & -1%
\end{array}%
\right) \left( 
\begin{array}{c}
\zeta _{\rho } \\ 
\zeta _{\eta } \\ 
\zeta _{\chi }%
\end{array}%
\right) .
\end{equation}

The charged scalar mass eigenstates are
\begin{equation}
\left( 
\begin{array}{c}
H_{1}^{\pm } \\ 
G_{1}^{\pm }%
\end{array}%
\right) =\left( 
\begin{array}{cc}
\cos \nu & \sin \nu \\ 
\sin \nu & -\cos \nu%
\end{array}%
\right) \left( 
\begin{array}{c}
\rho _{1}^{\pm } \\ 
\eta _{2}^{\pm }%
\end{array}%
\right) ,\hspace{1cm}\hspace{1cm}\left( 
\begin{array}{c}
H_{2}^{\pm } \\ 
G_{2}^{\pm }%
\end{array}%
\right) =\left( 
\begin{array}{cc}
\cos \gamma & \sin \gamma \\ 
\sin \gamma & -\cos \gamma%
\end{array}%
\right) \left( 
\begin{array}{c}
\rho _{3}^{\pm } \\ 
\chi _{2}^{\pm }%
\end{array}%
\right) .
\end{equation}

The remaining neutral scalar mass eigenstates are $\hspace{1cm}$
\begin{equation}
H_{2}^{0}\simeq \eta _{3}^{0},\hspace{1cm}\overline{H}_{2}^{0}\simeq 
\overline{\eta }_{3}^{0}\hspace{1cm}G_{2}^{0}\simeq -\chi _{1}^{0},\hspace{%
1cm}\overline{G}_{2}^{0}\simeq -\overline{\chi }_{1}^{0},
\end{equation}

The mixing angles of the physical scalar fields are
\begin{equation}
\tan \alpha \simeq \tan \beta \simeq \tan \nu \simeq \frac{v_{\rho }}{%
v_{\eta }},\hspace{1cm}\hspace{1cm}\tan \gamma \simeq \frac{v_{\rho }}{%
v_{\chi }}.
\end{equation}

Notice that after the spontaneous breaking of the gauge symmetry $%
SU(3)_{L}\otimes U(1)_{X}$ and rotations into mass eigenstates, the model
contains four massive charged Higgs ($H_{1}^{\pm }$, $H_{2}^{\pm }$), one
CP-odd Higgs ($A^{0}$), three neutral CP-even Higgs ($h^{0},H_{1}^{0},H_{3}^{0}$%
) and two neutral Higgs ($H_{2}^{0},\overline{H}_{2}^{0}$) bosons. Here we
identify the scalar $h^{0}$ with the SM-like $126$ GeV Higgs boson observed
at the LHC. We recall that the neutral Goldstone bosons $G_{1}^{0}$, $%
G_{3}^{0}$ correspond to the $Z$, $Z^{\prime }$\ gauge bosons, respectively,
while the remaining neutral Goldstone bosons $G_{2}^{0}$ , $\overline{G}%
_{2}^{0}$ correspond to the $K^{0}$, $\overline{K}^{0}$ gauge bosons,
respectively. Furthermore, the charged Goldstone bosons $G_{1}^{\pm }$ and $%
G_{2}^{\pm }$ correspond to the $W^{\pm }$ and $K^{\pm }$ gauge bosons,
respectively \cite{331-pisano,M-O}.

In Ref. \cite{Hernandez:2013hea} we follow the method described in Ref. \cite%
{Maniatis:2006fs}\ to show that the scalar potential is stable when its
quartic couplings satisfy the following relations: 
\begin{equation}
\lambda _{1}>0,\hspace{1cm}\lambda _{2}>0,\hspace{1cm}\lambda _{6}>0,\hspace{%
1cm}\lambda _{2}>\lambda _{3},\hspace{1cm}\lambda _{2}+\lambda _{4}>0,%
\hspace{1cm}\lambda _{5}+\lambda _{6}>2\sqrt{\lambda _{1}\left( \lambda
_{2}+\lambda _{4}\right) }.  \label{S7}
\end{equation}

\section{Lepton masses and mixing}

\label{lepton}

\subsection{Neutrino masses}

From Eq. (\ref{lyl}), and using the product rules for the $S_{3}$ group
given in Appendix \ref{appA}, it follows that the Yukawa mass terms for the
lepton sector are given by
\begin{eqnarray}
-\mathcal{L}_{mass}^{\left( L\right) } &=&\frac{v_{\rho }}{\sqrt{2}}%
\overline{e_{L}}h_{\rho e}^{\left( L\right) }e_{R}+\frac{v_{\eta }}{\sqrt{2}}%
\overline{\nu _{L}}h_{\Phi }^{\left( L\right) }N_{R}^{2}+\frac{v_{\chi }}{%
\sqrt{2}}\overline{\nu _{R}^{C}}h_{\Phi }^{\left( L\right) }N_{R}^{3}  \notag
\\
&&+\frac{1}{2}m_{N}^{\left( 1\right) }\overline{N}_{R}^{1}N_{R}^{1C}+\frac{1%
}{2}m_{N}\left( \overline{N}_{R}^{2}N_{R}^{2C}+\overline{N}%
_{R}^{3}N_{R}^{3C}\right)  \notag \\
&&+i\frac{\left( w_{\chi }v_{\eta }\sin \varphi _{\chi }-w_{\eta }v_{\chi
}\sin \varphi _{\eta }\right) v_{\rho }}{\sqrt{2}\Lambda ^{2}}\left( 
\overline{\nu }_{R}^{C}h_{\rho }\nu _{L}^{C}-\overline{\nu }_{L}h_{\rho }\nu
_{R}\right) +\text{H.c.}  \label{Llmass}
\end{eqnarray}%
We can rewrite the neutrino mass terms as 
\begin{equation}
-\mathcal{L}_{mass}^{\left( \nu \right) }=\frac{1}{2}\left( 
\begin{array}{ccc}
\overline{\nu _{L}^{C}} & \overline{\nu _{R}} & \overline{N_{R}}%
\end{array}%
\right) M_{\nu }\left( 
\begin{array}{c}
\nu _{L} \\ 
\nu _{R}^{C} \\ 
N_{R}^{C}%
\end{array}%
\right) +\text{H.c.},  \label{Lnu}
\end{equation}%
where the $S_{3}$ flavor symmetry constrains the neutrino mass matrix to be
of the form:%
\begin{equation}
M_{\nu }=\left( 
\begin{array}{ccc}
0_{3\times 3} & i\varepsilon v_{\rho } & Fv_{\eta } \\ 
i\varepsilon ^{T}v_{\rho } & 0_{3\times 3} & Gv_{\chi } \\ 
F^{T}v_{\eta } & G^{T}v_{\chi } & M_{R}%
\end{array}%
\right) ,  \label{Mnub}
\end{equation}%
with 
\begin{equation}
\varepsilon ^{\ast }=\frac{\left( h_{\rho }-h_{\rho }^{T}\right) \left(
w_{\chi }v_{\eta }\sin \varphi _{\chi }-w_{\eta }v_{\chi }\sin \varphi
_{\eta }\right) }{\sqrt{2}\Lambda ^{2}}=\left( 
\begin{array}{ccc}
0 & b_{3} & b_{2} \\ 
-b_{3} & 0 & b_{1} \\ 
-b_{2} & -b_{1} & 0%
\end{array}%
\right) ,  \label{b}
\end{equation}%
and the submatrices are defined by
\begin{equation}
F=\left( 
\begin{array}{ccc}
0 & a_{1} & 0 \\ 
0 & a_{2} & 0 \\ 
0 & a_{3} & 0%
\end{array}%
\right) ,\qquad G=\left( 
\begin{array}{ccc}
0 & 0 & a_{1} \\ 
0 & 0 & a_{2} \\ 
0 & 0 & a_{3}%
\end{array}%
\right) ,\qquad M_{R}=\left( 
\begin{array}{ccc}
\kappa m_{N} & 0 & 0 \\ 
0 & m_{N} & 0 \\ 
0 & 0 & m_{N}%
\end{array}%
\right) ,  \label{b3}
\end{equation}%
where $a_{j}=h_{\Phi j}^{\left( L\right) }/\sqrt{2}$ for$\ j=1,2,3$.

The ansatz for the matrices $F$, $G$ and $M_{R}$ follow from the fermion
assignments into $S_{3}$ irreducible representations, in particular the $%
S_{3}$ discrete symmetry constrains the heavy Majorana neutrino mass matrix $%
M_{R}$ to be diagonal.

%
%
%
%
%
%
%
%

\subsubsection{Diagonalization of the mass matrix}

Here, for simplicity we assume a scenario corresponding to a double seesaw
mechanism \cite{catano} where the heavy Majorana neutrino masses and the
VEV's satisfy the hierarchy 
\begin{equation}
\left( M_{R}\right) _{ll}\gg v_{\chi }\gg v_{\rho },v_{\eta }\gg w_{\chi
},w_{\eta },\qquad \qquad l=1,2,3.
\end{equation}

Resulting from this double seesaw mechanism we have three different mass
scales for the neutrinos: very light active neutrinos $\nu _{l}^{\left(
1\right) }$, light $\nu _{l}^{\left( 2\right) }$ and very heavy sterile
neutrinos $\nu _{l}^{\left( 3\right) }$ ($l=1,2,3$). As shown in detail in
Appendix \ref{appC}, their corresponding mass matrices satisfy the
relations: 
\begin{align}
M_{\nu }^{\left( 1\right) }\left( M_{\nu }^{\left( 1\right) }\right) ^{T}&
=xv_{\eta }^{2}\left( A-\frac{v_{\rho }^{2}}{xv_{\eta }^{2}}\varepsilon
\varepsilon ^{T}\right) ,  \label{M1} \\
M_{\nu }^{\left( 2\right) }\left( M_{\nu }^{\left( 2\right) }\right) ^{T}&
=zv_{\chi }^{2}\left( A-\frac{v_{\rho }^{2}}{zv_{\chi }^{2}}\varepsilon
^{T}\varepsilon \right) ,  \label{M2} \\
M_{\nu }^{\left( 3\right) }\left( M_{\nu }^{\left( 3\right) }\right) ^{T}&
=\left( 
\begin{array}{ccc}
\kappa ^{2}m_{N}^{2} & 0 & 0 \\ 
0 & m_{N}^{2} & 0 \\ 
0 & 0 & m_{N}^{2}%
\end{array}%
\right) ,  \label{M3}
\end{align}%
with 
\begin{equation}
x=\left( a_{1}^{2}+a_{2}^{2}+a_{3}^{2}\right) \frac{v_{\eta }^{2}}{m_{N}^{2}}%
,\qquad \qquad z=\left( a_{1}^{2}+a_{2}^{2}+a_{3}^{2}\right) \frac{v_{\chi
}^{2}}{m_{N}^{2}},  \label{xz}
\end{equation}%
where we assumed that the elements of the neutrino mass matrix of Eq. (\ref%
{Mmnu}) are real. Moreover, the active light neutrino mass matrix satisfies 
\begin{equation}
M_{\nu }^{\left( 1\right) }\left( M_{\nu }^{\left( 1\right) }\right)
^{T}\simeq xv_{\eta }^{2}\left( 
\begin{array}{ccc}
a_{1}^{2}+d_{2}^{2}+d_{3}^{2} & a_{1}a_{2}+d_{1}d_{2} & a_{1}a_{3}-d_{1}d_{3}
\\ 
a_{1}a_{2}+d_{1}d_{2} & a_{2}^{2}+d_{1}^{2}+d_{3}^{2} & a_{2}a_{3}+d_{2}d_{3}
\\ 
a_{1}a_{3}-d_{1}d_{3} & a_{2}a_{3}+d_{2}d_{3} & a_{3}^{2}+d_{1}^{2}+d_{2}^{2}%
\end{array}%
\right) ,
\end{equation}%
where%
\begin{equation}
d_{j}=i\frac{v_{\rho }}{\sqrt{x}v_{\eta }}b_{j},\qquad \qquad j=1,2,3,
\label{d}
\end{equation}%
and $b_{j}$ are purely imaginary.

The squared light neutrino mass matrix $M_{\nu }^{\left( 1\right) }\left(
M_{\nu }^{\left( 1\right) }\right) ^{T}$ is diagonalized by a rotation
matrix $R_{\nu }$, according to: 
\begin{equation}
\allowbreak R_{\nu }^{T}M_{\nu }^{\left( 1\right) }\left( M_{\nu }^{\left(
1\right) }\right) ^{T}R_{\nu }=\left( 
\begin{array}{ccc}
m_{1}^{2} & 0 & 0 \\ 
0 & m_{2}^{2} & 0 \\ 
0 & 0 & m_{3}^{2}%
\end{array}%
\right) ,\qquad \qquad R_{\nu }=\left( 
\begin{array}{ccc}
-\cos \xi _{1}\sin \xi _{2} & -\sin \xi _{1} & \cos \xi _{1}\cos \xi _{2} \\ 
\cos \xi _{2} & 0 & \sin \xi _{2} \\ 
\sin \xi _{1}\sin \xi _{2} & \cos \xi _{1} & \cos \xi _{2}\sin \xi _{1}%
\end{array}%
\right) \allowbreak ,  \label{Rnu}
\end{equation}%
where 
\begin{equation}
\tan \xi _{1}\simeq \frac{a_{3}}{a_{1}},\qquad \qquad \tan 2\xi _{2}\simeq 
\frac{2a_{2}\sqrt{a_{1}^{2}+a_{3}^{2}}}{\left(
a_{1}^{2}+a_{3}^{2}-a_{2}^{2}\right) }.
\end{equation}%
Here we have also assumed that: 
\begin{equation}
\sigma _{j}=\frac{d_{j}}{a_{j}}=\frac{\sqrt{2}v_{\rho }b_{j}}{\sqrt{x}%
v_{\eta }h_{\Phi j}^{\left( L\right) }}=\sigma ,\qquad j=1,2,3.
\label{sigma}
\end{equation}%
The squared light neutrino masses are given by
\begin{eqnarray}
m_{1}^{2} &\simeq &\sigma ^{2}\left[ a_{1}^{2}+a_{2}^{2}+a_{3}^{2}-\frac{%
4a_{2}^{2}\left( a_{1}^{2}+a_{3}^{2}\right) }{\left(
a_{1}^{2}+a_{2}^{2}+a_{3}^{2}\right) }\right] xv_{\eta }^{2},  \notag \\
m_{2}^{2} &\simeq &\sigma ^{2}\left( a_{1}^{2}+a_{2}^{2}+a_{3}^{2}\right)
xv_{\eta }^{2},  \notag \\
m_{3}^{2} &\simeq &\left( a_{1}^{2}+a_{2}^{2}+a_{3}^{2}\right) xv_{\eta
}^{2}.  \label{numasses}
\end{eqnarray}%
We thus predict a normal hierarchy neutrino mass spectrum, with neutrino
mass squared splittings 
\begin{equation}
\Delta m_{21}^{2}\simeq \frac{4\sigma ^{2}a_{2}^{2}\left(
a_{1}^{2}+a_{3}^{2}\right) xv_{\eta }^{2}}{\left(
a_{1}^{2}+a_{2}^{2}+a_{3}^{2}\right) },\qquad \qquad \Delta m_{31}^{2}\simeq
\left( a_{1}^{2}+a_{2}^{2}+a_{3}^{2}\right) xv_{\eta }^{2}.
\label{Deltamsquared}
\end{equation}

Notice that both the six-dimensional Yukawa term in Eq. (\ref{lyl}) as well
as the nonvanishing vacuum expectation values in the first and third
components of the $\chi $ and $\eta $ triplets, respectively, are crucial to
get a nonzero solar neutrino mass squared splitting $\Delta m_{21}^{2}$
since they correspond to $\sigma \neq 0$. Furthermore, the hierarchy $\Delta
m_{21}^{2}\ll \Delta m_{31}^{2}$ can be explained as a consequence of the
subleading contribution to the neutrino mass matrix arising from the six
dimensional Yukawa term in Eq. (\ref{lyl}) proportional to $b_{j}$.

The orders of magnitude of the SM particles and new physics give the initial
constraints $v_{\chi }\gtrsim 1$ TeV and $v_{\eta }^{2}+v_{\rho }^{2}=v^{2}$%
. We can choose to set the Yukawa couplings $h_{\Phi 1}^{\left( L\right)
}\sim h_{\Phi 2}^{\left( L\right) }\sim h_{\Phi 3}^{\left( L\right) }$,
following the definition (\ref{b3}) implies $a_{j}\sim a$ ($j=1,2,3$). We
also assume that $b_{j}\sim b$ ($j=1,2,3$), which means that the elements of
the Yukawa matrix $h_{\rho }$ are of the same order. From Eq. (\ref{xz}), $%
x\sim a^{2}\frac{v_{\eta }^{2}}{m_{N}^{2}}$ and in first approximation $%
\Delta m_{31}^{2}\sim \frac{a^{4}v_{\eta }^{4}}{m_{N}^{2}}$ [see Eq. (\ref%
{Deltamsquared})]. Therefore, in order to get the right order of magnitude
of the atmospheric neutrino mass squared splitting $\Delta m_{31}^{2}$, we
need the heavy neutrinos $N_{R}^{2,3}$ to have mass $m_{N}\sim 10^{14}a^{2}$
GeV. In addition, we also get the estimate $\frac{\Delta m_{31}^{2}}{\Delta
m_{21}^{2}}$ $\sim \frac{1-\sigma ^{2}}{\sigma ^{2}}$, and since the
experimental data on neutrino oscillations implies $\frac{\Delta m_{31}^{2}}{%
\Delta m_{21}^{2}}$ $\sim 30$, this gives $\sigma \sim 10^{-1}$, which
results, according to Eq. (\ref{sigma}), in $d\sim 10^{-1}a$. From Eq. (\ref%
{Deltamsquared}), we get that our estimate $d\sim 10^{-1}a$ yields $%
a^{2}x\sim 10^{-25}$, which implies, according to Eq. (\ref{d}), that $%
\left\vert b\right\vert \sim 10^{-14}$. Furthermore, from Eq. (\ref{numasses}%
) we get for the light active neutrino masses the estimates $m_{1}\sim
m_{2}\sim 6$ meV and $m_{3}\sim 30$ meV, which corresponds to a
quasidegenerate normal hierarchy neutrino mass spectrum. Besides that, we
get that the heaviest sterile neutrino has a mass of about $M_{3}\sim 3$
keV. Assuming $\left\vert \left( h_{\rho }\right) _{jl}\right\vert \sim 1$ ($%
j,l=1,2,3$), taking into account $\left\vert b\right\vert \sim 10^{-14}$ and
using Eq. (\ref{b}) and considering $v_{\chi }\sim 1$ TeV and $w_{\chi }\sim
w_{\eta }\sim 1$ GeV we get for the cutoff of our model the estimate 
\begin{equation}
\Lambda \sim 10^{4}-10^{5}\text{ TeV}.  \label{cutoff}
\end{equation}

\subsection{Charged leptons}

Regarding the charged leptons, we assume that the corresponding mass matrix
is that one given by the Fukuyama-Nishiura ansatz \cite{textures}, as
follows: 
\begin{equation}
M_{l}=\frac{v_{\rho }}{\sqrt{2}}h_{\rho e}^{\left( L\right) }=\frac{v_{\rho }%
}{\sqrt{2}}\left( 
\begin{array}{ccc}
0 & h_{1}e^{i\gamma } & h_{1}e^{i\gamma } \\ 
h_{1}e^{-i\gamma } & h_{2} & h_{3} \\ 
h_{1}e^{-i\gamma } & h_{3} & h_{2}%
\end{array}%
\right) \allowbreak \allowbreak =P_{l}\widetilde{M}_{l}P_{l}^{\dagger
},\allowbreak
\end{equation}%
where 
\begin{equation}
\widetilde{M}_{l}=\frac{v_{\rho }}{\sqrt{2}}\left( 
\begin{array}{ccc}
0 & h_{1} & h_{1} \\ 
h_{1} & h_{2} & h_{3} \\ 
h_{1} & h_{3} & h_{2}%
\end{array}%
\right) \allowbreak ,\qquad \qquad P_{l}=\left( 
\begin{array}{ccc}
1 & 0 & 0 \\ 
0 & e^{-i\gamma } & 0 \\ 
0 & 0 & e^{-i\gamma }%
\end{array}%
\right) .
\end{equation}

Then, the charged lepton masses are given by
\begin{eqnarray}
-m_{e} &=&\frac{v_{\rho }}{2\sqrt{2}}\left( h_{2}+h_{3}-\sqrt{\left(
h_{2}-h_{3}\right) ^{2}+8h_{1}^{2}}\right) , \\
m_{\mu } &=&\frac{v_{\rho }}{2\sqrt{2}}\left( h_{2}+h_{3}+\sqrt{\left(
h_{2}-h_{3}\right) ^{2}+8h_{1}^{2}}\right) , \\
m_{\tau } &=&\frac{v_{\rho }}{\sqrt{2}}\left( h_{2}-h_{3}\right) ,
\end{eqnarray}%
and the mass matrix $\widetilde{M}_{l}$ is diagonalized by a rotation matrix 
$\widetilde{R}_{l}$ according to
\begin{equation}
\widetilde{R}_{l}^{T}\widetilde{M}_{l}\widetilde{R}_{l}=\left( 
\begin{array}{ccc}
-m_{e} & 0 & 0 \\ 
0 & m_{\mu } & 0 \\ 
0 & 0 & m_{\tau }%
\end{array}%
\right) ,\qquad \qquad \widetilde{R}_{l}=\left( 
\begin{array}{ccc}
c_{l} & s_{l} & 0 \\ 
-\frac{s_{l}}{\sqrt{2}} & \frac{c_{l}}{\sqrt{2}} & -\frac{1}{\sqrt{2}} \\ 
-\frac{s_{l}}{\sqrt{2}} & \frac{c_{l}}{\sqrt{2}} & \frac{1}{\sqrt{2}}%
\end{array}%
\right) ,
\end{equation}%
with 
\begin{equation}
c_{l}=\cos \theta _{l}=\sqrt{\frac{m_{\mu }}{m_{\mu }+m_{e}}},\qquad \qquad
s_{l}=\sin \theta _{l}=\sqrt{\frac{m_{e}}{m_{\mu }+m_{e}}}.  \label{thetal}
\end{equation}

Putting it all together, the charged lepton mass matrix $M_{l}$ is
diagonalized by a rotation matrix $R_{l}$ according to
\begin{equation}
R_{l}^{\dagger }M_{l}R_{l}=\left( 
\begin{array}{ccc}
-m_{e} & 0 & 0 \\ 
0 & m_{\mu } & 0 \\ 
0 & 0 & m_{\tau }%
\end{array}%
\right) ,\qquad \qquad R_{l}=P_{l}\widetilde{R}_{l}.  \label{Rl}
\end{equation}

\subsection{Lepton mixing}

With the rotation matrices in the charged lepton sector $R_{l}$ given in Eq.
(\ref{Rl}), and in the neutrino sector $R_{\nu }$ given in Eq. (\ref{Rnu}),
we find the PMNS leptonic mixing matrix
\begin{equation}
U=\left( 
\begin{array}{ccc}
-c_{1}s_{2}c_{l}-\frac{1}{\sqrt{2}}c_{2}s_{l}e^{i\gamma }-\frac{1}{\sqrt{2}}%
s_{1}s_{2}s_{l}e^{i\gamma } & -s_{1}c_{l}-\frac{1}{\sqrt{2}}%
c_{1}s_{l}e^{i\gamma } & c_{1}c_{2}c_{l}-\frac{1}{\sqrt{2}}%
s_{2}s_{l}e^{i\gamma }-\frac{1}{\sqrt{2}}c_{2}s_{1}s_{l}e^{i\gamma } \\ 
\frac{1}{\sqrt{2}}c_{2}c_{l}e^{i\gamma }-c_{1}s_{2}s_{l}+\frac{1}{\sqrt{2}}%
s_{1}s_{2}c_{l}e^{i\gamma } & \frac{1}{\sqrt{2}}c_{1}c_{l}e^{i\gamma
}-s_{1}s_{l} & c_{1}c_{2}s_{l}+\frac{1}{\sqrt{2}}s_{2}c_{l}e^{i\gamma }+%
\frac{1}{\sqrt{2}}c_{2}s_{1}c_{l}e^{i\gamma } \\ 
\frac{1}{\sqrt{2}}s_{1}s_{2}e^{i\gamma }-\frac{1}{\sqrt{2}}c_{2}e^{i\gamma }
& \frac{1}{\sqrt{2}}c_{1}e^{i\gamma } & \frac{1}{\sqrt{2}}%
c_{2}s_{1}e^{i\gamma }-\frac{1}{\sqrt{2}}s_{2}e^{i\gamma }%
\end{array}%
\right) \allowbreak . \label{UPMNS}
\end{equation}

From the standard parametrization of the lepton mixing matrix, it follows
that the lepton mixing angles are
\begin{eqnarray}
\sin ^{2}\theta _{12} &=&\frac{\left\vert U_{e2}\right\vert ^{2}}{%
1-\left\vert U_{e3}\right\vert ^{2}} \\
&=&\frac{\frac{1}{2}\sin ^{2}\gamma \sin ^{2}\theta _{l}\cos ^{2}\xi _{1}+%
\frac{1}{4}\left( -\sqrt{2}\cos \gamma \sin \theta _{l}\cos \xi _{1}-2\cos
\theta _{l}\sin \xi _{1}\right) ^{2}}{1-\frac{1}{2}\sin ^{2}\gamma \sin
^{2}\theta _{l}\left( \sin \xi _{2}+\sin \xi _{1}\cos \xi _{2}\right) {}^{2}-%
\frac{1}{4}\left[ 2\cos \theta _{l}\cos \xi _{1}\cos \xi _{2}-\sqrt{2}\cos
\gamma \sin \theta _{l}\left( \sin \xi _{2}+\sin \xi _{1}\cos \xi
_{2}\right) \right] ^{2}},  \notag \\
\end{eqnarray}
\begin{eqnarray}
\sin ^{2}\theta _{13} &=&\left\vert U_{e3}\right\vert ^{2} \\
&=&\frac{1}{2}\sin ^{2}\gamma \sin ^{2}\theta _{l}\left( \sin \xi _{2}+\sin
\xi _{1}\cos \xi _{2}\right) {}^{2}+\frac{1}{4}\left[ 2\cos \theta _{l}\cos
\xi _{1}\cos \xi _{2}-\sqrt{2}\cos \gamma \sin \theta _{l}\left( \sin \xi
_{2}+\sin \xi _{1}\cos \xi _{2}\right) \right] ^{2},  \notag \\
\end{eqnarray}
\begin{eqnarray}
\sin ^{2}\theta _{23} &=&\frac{\left\vert U_{\mu 3}\right\vert ^{2}}{%
1-\left\vert U_{e3}\right\vert ^{2}} \\
&=&\frac{\frac{1}{2}\sin ^{2}\gamma \cos ^{2}\theta _{l}\left( \sin \xi
_{2}+\sin \xi _{1}\cos \xi _{2}\right) {}^{2}+\frac{1}{4}\left( \sqrt{2}\cos
\gamma \cos \theta _{l}\left( \sin \xi _{2}+\sin \xi _{1}\cos \xi
_{2}\right) +2\sin \theta _{l}\cos \xi _{1}\cos \xi _{2}\right) {}^{2}}{1-%
\frac{1}{2}\sin ^{2}\gamma \sin ^{2}\theta _{l}\left( \sin \xi _{2}+\sin \xi
_{1}\cos \xi _{2}\right) {}^{2}-\frac{1}{4}\left[ 2\cos \theta _{l}\cos \xi
_{1}\cos \xi _{2}-\sqrt{2}\cos \gamma \sin \theta _{l}\left( \sin \xi
_{2}+\sin \xi _{1}\cos \xi _{2}\right) \right] ^{2}}.  \notag \\
\end{eqnarray}

The Jarlskog invariant and the CP violating phase are respectively given by 
\cite{PDG,Jarlskog,Branco}%
\begin{equation}
J=\func{Im}\left( U_{e1}U_{\mu 2}U_{e2}^{\ast }U_{\mu 1}^{\ast }\right) =-%
\frac{\sin \gamma \sin 2\theta _{l}}{32\sqrt{2}}\left[ 4\cos ^{2}\xi
_{2}\sin 2\xi _{1}-6\sin 4\xi _{1}\sin ^{2}\xi _{2}+\sin 2\xi _{2}\left(
\cos \xi _{1}-5\cos 3\xi _{1}\right) \right] ,
\end{equation}%
\begin{equation}
\sin \delta =\frac{\left( 1-\bigl|U_{e3}\bigr|^{2}\right) J}{\bigl|%
U_{e1}U_{e2}U_{e3}U_{\mu 3}U_{\tau 3}\bigr|}\,.
\end{equation}

Varying the parameters $\xi _{1}$, $\xi _{2}$ and $\gamma $ we have fitted
the $\sin ^{2}\theta _{ij}$ to the experimental values in Table \ref{NH} for
the normal hierarchy neutrino mass spectrum. From Eq. (\ref{thetal}), we
have
\begin{equation}
\sin 2\theta _{l}\simeq 2\sqrt{\frac{m_{e}}{m_{\mu }}}.
\end{equation}

The best fit result is
\begin{equation}
\xi _{1}=212.8^{\circ },\qquad \xi _{2}=101.7^{\circ },\qquad \gamma
=66.4^{\circ }.
\end{equation}%
\begin{eqnarray}
&&\sin ^{2}\theta _{12}=0.32,\qquad \qquad \sin ^{2}\theta
_{23}=0.613,\qquad \qquad \sin ^{2}\theta _{13}=0.0246,
\label{Parameter-fit-NH} \\[0.12in]
&&J=-9.81\times 10^{-3},\qquad \qquad \delta =16^{\circ }.  \notag
\end{eqnarray}%
Comparing the results in (\ref{Parameter-fit-NH}) with the values in Table %
\ref{NH}, we see that the mixing parameters $\sin ^{2}\theta _{12}$, $\sin
^{2}\theta _{13}$ and $\sin ^{2}\theta _{23}$ are in agreement with the
experimental data. We obtain that CP is violated in neutrino oscillations
with a Jarlskog invariant of about $10^{-2}$. Furthermore, the complex phase 
$\gamma $ responsible for CP violation in lepton sector arises from the
Yukawa terms for the charged leptons. 
\begin{table}[tbh]
\begin{tabular}{|c|c|c|c|c|c|}
\hline
Parameter & $\Delta m_{21}^{2}$($10^{-5}$eV$^2$) & $\Delta m_{31}^{2}$($%
10^{-3}$eV$^2$) & $\left( \sin ^{2}\theta _{12}\right) _{\exp }$ & $\left(
\sin ^{2}\theta _{23}\right) _{\exp }$ & $\left( \sin ^{2}\theta
_{13}\right) _{\exp }$ \\ \hline
Best fit & $7.62$ & $2.55$ & $0.320$ & $0.613$ & $0.0246$ \\ \hline
$1\sigma $ range & $7.43-7.81$ & $2.46-2.61$ & $0.303-0.336$ & $0.573-0.635$
& $0.0218-0.0275$ \\ \hline
$2\sigma $ range & $7.27-8.01$ & $2.38-2.68$ & $0.29-0.35$ & $0.38-0.66$ & $%
0.019-0.030$ \\ \hline
$3\sigma $ range & $7.12-8.20$ & $2.31-2.74$ & $0.27-0.37$ & $0.36-0.68$ & 
\\ \hline
\end{tabular}%
\caption{Range for experimental values of neutrino mass squared splittings
and leptonic mixing parameters taken from Ref. \protect\cite{Tortola:2012te}
for the case of normal hierarchy.}
\label{NH}
\end{table}

\section{Conclusions}

\label{conclusions}

We proposed a model based on the group $SU(3)_{C}\otimes SU(3)_{L}\otimes
U(1)_{X}\otimes S_{3}$ where lepton masses and mixing can be reproduced. We
assumed that the heavy Majorana neutrinos have masses much larger than the
TeV scale, so that the hierarchy $m_{N}\gg v_{\chi }\gg v_{\rho },v_{\eta }$
is fulfilled, implying that the small active neutrino masses are generated
via a double seesaw mechanism. We found that the only $SU\left( 3\right)
_{L}\otimes U\left( 1\right) _{X}\otimes S_{3}$ invariant nonrenormalizable
operators of lowest order that contribute to the neutrino masses are $\frac{1%
}{\Lambda }\overline{N}_{R}^{1}N_{R}^{C}\left( \Phi \Phi ^{\dagger }\right) $
and $\frac{1}{\Lambda ^{2}}\overline{L}_{L}L_{L}^{C}\rho \left( \Phi
^{\dagger }\Phi \right) $. From these nonrenormalizable terms, $\frac{1}{%
\Lambda ^{2}}\overline{L}_{L}L_{L}^{C}\rho \left( \Phi ^{\dagger }\Phi
\right) $ gives a relevant contribution to the neutrino masses since the
operator $\frac{1}{\Lambda }\overline{N}_{R}^{1}N_{R}^{C}\left( \Phi \Phi
^{\dagger }\right) $ gives a subleading contribution to the heavy Majorana
neutrino masses.

In this scenario, the spectrum of neutrinos presents very light, light and
very heavy masses. Assuming that the heavy Majorana neutrinos have masses of
about $m_{N}\sim 10^{14}$ GeV, we find for the light active neutrino masses
the estimates $m_{1}\sim m_{2}\sim 6$ meV and $m_{3}\sim 30$ meV, while for
the heaviest sterile neutrino mass we get $M_{3}\sim 3$ keV.\ The model
predicts a quasidegenerate normal hierarchy active neutrino mass spectrum
and the relation $\Delta m_{21}^{2}\ll \Delta m_{31}^{2}$ can be explained
as consequence of the small mass terms arising from the effective six-
dimensional operators. We find for the scale of these effective operators
the estimate $\Lambda \sim 10^{4}-10^{5}$ TeV. These effective operators
generate a nonvanishing solar neutrino mass squared splitting $\Delta
m_{21}^{2}$, provided that the first and third components of the $\chi $ and 
$\eta $ triplets, respectively, should have nonvanishing vacuum expectation
values, where at least one of them has to be complex. The obtained neutrino
mixing parameters are in excellent agreement with the neutrino oscillation
experimental data. We find that CP is violated in neutrino oscillations with
a Jarlskog invariant of about $10^{-2}$. Furthermore, the complex phase
responsible for CP violation in the lepton sector has been assumed to come
from the Yukawa terms for the charged leptons.

\section*{Acknowledgments}

A.E.C.H was supported by Fondecyt (Chile), Grant No. 11130115 and by DGIP
internal Grant No. 111458. R.M. was supported by COLCIENCIAS and by Fondecyt
(Chile), Grant No. 11130115. \appendix

\section{The product rules for $S_{3}$}

\label{appA}

The $S_{3}$ group has three irreducible representations: $\mathbf{1}$, $%
\mathbf{1}^{\prime }$ and $\mathbf{2}$. Denoting $\left( x_{1},x_{2}\right)
^{T}$\ and $\left( y_{1},y_{2}\right) ^{T}$ as the basis vectors for two $%
S_{3}$ doublets and $y%
{\acute{}}%
$ a nontrivial $S_{3}$ singlet, the multiplication rules of the $S_{3}$
group for the case of real representations are given by \cite%
{Ishimori:2010au}
\begin{equation}
\left( 
\begin{array}{c}
x_{1} \\ 
x_{2}%
\end{array}%
\right) _{\mathbf{2}}\otimes \left( 
\begin{array}{c}
y_{1} \\ 
y_{2}%
\end{array}%
\right) _{\mathbf{2}}=\left( x_{1}y_{1}+x_{2}y_{2}\right) _{\mathbf{1}%
}+\left( x_{1}y_{2}-x_{2}y_{1}\right) _{\mathbf{1}^{\prime }}+\left( 
\begin{array}{c}
x_{1}y_{2}+x_{2}y_{1} \\ 
x_{1}y_{1}-x_{2}y_{2}%
\end{array}%
\right) _{\mathbf{2}},  \label{6}
\end{equation}%
\begin{equation}
\left( 
\begin{array}{c}
x_{1} \\ 
x_{2}%
\end{array}%
\right) _{\mathbf{2}}\otimes \left( y\right) _{\mathbf{1}^{\prime }}=\left( 
\begin{array}{c}
-x_{2}y \\ 
x_{1}y%
\end{array}%
\right) _{\mathbf{2}},\qquad \qquad \left( x\right) _{\mathbf{1}^{\prime
}}\otimes \left( y\right) _{\mathbf{1}^{\prime }}=\left( xy\right) _{\mathbf{%
1}}.  \label{7}
\end{equation}

\section{Diagonalization of the neutrino mass matrix}

\label{appC} We consider the neutrino mass matrix 
\begin{equation}
M_{\nu }=\left( 
\begin{array}{ccc}
0_{3\times 3} & i\varepsilon v_{\rho } & Fv_{\eta } \\ 
i\varepsilon ^{T}v_{\rho } & 0_{3\times 3} & Gv_{\chi } \\ 
F^{T}v_{\eta } & G^{T}v_{\chi } & M_{R}%
\end{array}%
\right) =\left( 
\begin{array}{cc}
\widetilde{M}_{\nu } & C \\ 
C^{T} & M_{R}%
\end{array}%
\right) ,  \label{Mmnu}
\end{equation}%
where the different sub-blocks are given by Eqs. (\ref{b})-(\ref{b3}).%

For the sake of simplicity we assume that the elements of the neutrino mass
matrix of Eq. (\ref{Mmnu}) are real. Now, in order to block-diagonalize the
mass matrix $M_{\nu }$, we apply the transformation
\begin{equation}
W^{T}M_{\nu }W\simeq \left( 
\begin{array}{cc}
\widetilde{M}_{\nu }-CB^{T}-BC^{T}+BM_{R}B^{T} & C-BM_{R} \\ 
C^{T}-M_{R}B^{T} & M_{R}+C^{T}B+B^{T}C%
\end{array}%
\right) ,  \notag
\end{equation}%
with 
\begin{equation}
W=\left( 
\begin{array}{cc}
1_{6\times 6} & B \\ 
-B^{T} & 1_{3\times 3}%
\end{array}%
\right) .  \label{C7}
\end{equation}%
Using the method of recursive expansion of Ref. \cite{grimus}, we find that
the block diagonalization condition leads to 
\begin{equation}
B\simeq CM_{R}^{-1},\qquad \qquad B^{T}\simeq M_{R}^{-1}C^{T}.  \label{C8}
\end{equation}%
Then, it follows that:%
\begin{equation}
W^{T}M_{\nu }W\simeq \left( 
\begin{array}{cc}
\widetilde{M}_{\nu }-CM_{R}^{-1}C^{T} & 0_{6\times 3} \\ 
0_{3\times 6} & M_{R}%
\end{array}%
\right) ,\qquad \qquad CM_{R}^{-1}C^{T}=\left( 
\begin{array}{cc}
FM_{R}^{-1}F^{T}v_{\eta }^{2} & FM_{R}^{-1}G^{T}v_{\eta }v_{\chi } \\ 
GM_{R}^{-1}F^{T}v_{\eta }v_{\chi } & GM_{R}^{-1}G^{T}v_{\chi }^{2}%
\end{array}%
\right) .  \label{C9}
\end{equation}%
The previous relations imply: 
\begin{equation}
W^{T}M_{\nu }W\simeq \left( 
\begin{array}{ccc}
-\frac{v_{\eta }^{2}}{m_{N}}A & i\varepsilon v_{\rho } & 0_{3\times 3} \\ 
i\varepsilon ^{T}v_{\rho } & -\frac{v_{\chi }^{2}}{m_{N}}A & 0_{3\times 3}
\\ 
0_{3\times 3} & 0_{3\times 3} & M_{R}%
\end{array}%
\right) =\left( 
\begin{array}{cc}
S_{\nu } & 0_{6\times 3} \\ 
0_{3\times 6} & M_{R}%
\end{array}%
\right) ,  \label{C10}
\end{equation}%
\bigskip where%
\begin{equation}
S_{\nu }=\left( 
\begin{array}{cc}
-\frac{v_{\eta }^{2}}{m_{N}}A & i\varepsilon v_{\rho } \\ 
i\varepsilon ^{T}v_{\rho } & -\frac{v_{\chi }^{2}}{m_{N}}A%
\end{array}%
\right) .  \label{C11}
\end{equation}%
Moreover, the matrix $S_{\nu }S_{\nu }^{T}$ gives 
\begin{equation}
S_{\nu }S_{\nu }^{T}=\left( 
\begin{array}{cc}
X & Y \\ 
Y^{T} & Z%
\end{array}%
\right) ,  \label{C18}
\end{equation}%
where 
\begin{equation}
X=xv_{\eta }^{2}A-v_{\rho }^{2}\varepsilon \varepsilon ^{T},\qquad
Z=zv_{\chi }^{2}A-v_{\rho }^{2}\varepsilon ^{T}\varepsilon ,\qquad
Y=iyv_{\rho }v_{\chi }\varepsilon A,  \label{C19}
\end{equation}%
with 
\begin{eqnarray}
x &=&\left( a_{1}^{2}+a_{2}^{2}+a_{3}^{2}\right) \frac{v_{\eta }^{2}}{%
m_{N}^{2}},\hspace{1cm}y=-\frac{v_{\chi }}{m_{N}},  \label{C0} \\
z &=&\left( a_{1}^{2}+a_{2}^{2}+a_{3}^{2}\right) \frac{v_{\chi }^{2}}{%
m_{N}^{2}}.  \notag
\end{eqnarray}%
Furthermore, the following hierarchy is fullfilled: 
\begin{equation}
\left\vert Y_{ij}\right\vert \ll \left\vert X_{ij}\right\vert \ll \left\vert
Z_{ij}\right\vert .  \label{C22}
\end{equation}

Now, in order to block-diagonalize the matrix $S_{\nu }S_{\nu }^{T}$, we
apply the transformation 
\begin{equation}
P^{T}S_{\nu }S_{\nu }^{T}P\simeq \left( 
\begin{array}{cc}
X-YK^{T}-KY^{T}+KZK^{T} & Y-KZ \\ 
Y^{T}-ZK^{T} & Z+Y^{T}K+K^{T}Y+K^{T}XK%
\end{array}%
\right) ,  \notag
\end{equation}%
with 
\begin{equation}
P=\left( 
\begin{array}{cc}
1_{3\times 3} & K \\ 
-K^{T} & 1_{3\times 3}%
\end{array}%
\right) .  \label{C24}
\end{equation}%
The block diagonalization condition leads to the relations 
\begin{equation}
K\simeq YZ^{-1},\qquad \qquad K^{T}=Z^{-1}Y^{T}.  \label{C25}
\end{equation}%
Therefore, we obtain 
\begin{equation}
P^{T}S_{\nu }S_{\nu }^{T}P\simeq \left( 
\begin{array}{cc}
M_{\nu }^{\left( 1\right) }\left( M_{\nu }^{\left( 1\right) }\right) ^{T} & 
0_{3\times 3} \\ 
0_{3\times 3} & M_{\nu }^{\left( 2\right) }\left( M_{\nu }^{\left( 2\right)
}\right) ^{T}%
\end{array}%
\right) ,  \label{C26}
\end{equation}%
where 
\begin{equation}
M_{\nu }^{\left( 1\right) }\left( M_{\nu }^{\left( 1\right) }\right)
^{T}\simeq xv_{\eta }^{2}\left( A-\frac{v_{\rho }^{2}}{xv_{\eta }^{2}}%
\varepsilon \varepsilon ^{T}\right) ,  \label{C27}
\end{equation}%
\begin{equation}
M_{\nu }^{\left( 2\right) }\left( M_{\nu }^{\left( 2\right) }\right)
^{T}\simeq zv_{\chi }^{2}\left( A-\frac{v_{\rho }^{2}}{zv_{\chi }^{2}}%
\varepsilon ^{T}\varepsilon \right) .  \label{C28}
\end{equation}%
Notice that $M_{\nu }^{\left( 1\right) }\left( M_{\nu }^{\left( 1\right)
}\right) ^{T}$ corresponds to the squared active light neutrino mass matrix.
Moreover, Eq. (\ref{C27}) can be rewritten as follows: 
\begin{equation}
M_{\nu }^{\left( 1\right) }\left( M_{\nu }^{\left( 1\right) }\right)
^{T}\simeq xv_{\eta }^{2}\left( 
\begin{array}{ccc}
a_{1}^{2}+d^{2}-d_{1}^{2} & a_{1}a_{2}+d_{1}d_{2} & a_{1}a_{3}-d_{1}d_{3} \\ 
a_{1}a_{2}+d_{1}d_{2} & a_{2}^{2}+d^{2}-d_{2}^{2} & a_{2}a_{3}+d_{2}d_{3} \\ 
a_{1}a_{3}-d_{1}d_{3} & a_{2}a_{3}+d_{2}d_{3} & a_{3}^{2}+d^{2}-d_{3}^{2}%
\end{array}%
\right) ,
\end{equation}%
where 
\begin{equation}
\tan 2\xi _{1}=\frac{2\left( a_{1}a_{3}-d_{1}d_{3}\right) }{%
a_{1}^{2}-a_{3}^{2}-d_{1}^{2}+d_{3}^{2}},\qquad \qquad \tan 2\xi _{2}=\frac{%
2a_{2}\left( 1+\sigma ^{2}\right) \sqrt{a_{1}^{2}+a_{3}^{2}}}{\left(
1-\sigma ^{2}\right) \left( a_{1}^{2}+a_{3}^{2}-a_{2}^{2}\right) },
\end{equation}%
so that the squared neutrino masses are given by
\begin{eqnarray}
m_{1}^{2} &=&\frac{xv_{\eta }^{2}}{2}\left( 1+\sigma ^{2}\right) \left(
a_{1}^{2}+a_{2}^{2}+a_{3}^{2}\right) -\frac{xv_{\eta }^{2}}{2}\sqrt{\left(
1+\sigma ^{4}\right) \left( a_{1}^{2}+a_{2}^{2}+a_{3}^{2}\right)
^{2}-2\sigma ^{2}\left[ \left( a_{1}^{2}+a_{3}^{2}+a_{2}^{2}\right)
^{2}-8a_{2}^{2}\left( a_{1}^{2}+a_{3}^{2}\right) \right] },  \notag \\
m_{2}^{2} &=&x\sigma ^{2}\left( a_{1}^{2}+a_{2}^{2}+a_{3}^{2}\right) v_{\eta
}^{2}, \\
m_{3}^{2} &=&\frac{xv_{\eta }^{2}}{2}\left( 1+\sigma ^{2}\right) \left(
a_{1}^{2}+a_{2}^{2}+a_{3}^{2}\right) +\frac{xv_{\eta }^{2}}{2}\sqrt{\left(
1+\sigma ^{4}\right) \left( a_{1}^{2}+a_{2}^{2}+a_{3}^{2}\right)
^{2}-2\sigma ^{2}\left[ \left( a_{1}^{2}+a_{3}^{2}+a_{2}^{2}\right)
^{2}-8a_{2}^{2}\left( a_{1}^{2}+a_{3}^{2}\right) \right] }.  \notag
\end{eqnarray}

Thus, the squared light neutrino mass matrix $M_{\nu }^{\left( 1\right)
}\left( M_{\nu }^{\left( 1\right) }\right) ^{T}$ is diagonalized by a
rotation matrix $R_{\nu }$, according to 
\begin{equation}
\allowbreak R_{\nu }^{T}M_{\nu }^{\left( 1\right) }\left( M_{\nu }^{\left(
1\right) }\right) ^{T}R_{\nu }=\left( 
\begin{array}{ccc}
m_{1}^{2} & 0 & 0 \\ 
0 & m_{2}^{2} & 0 \\ 
0 & 0 & m_{3}^{2}%
\end{array}%
\right) ,
\end{equation}%
where 
\begin{equation}
R_{\nu }=\left( 
\begin{array}{ccc}
-\cos \xi _{1}\sin \xi _{2} & -\sin \xi _{1} & \cos \xi _{1}\cos \xi _{2} \\ 
\cos \xi _{2} & 0 & \sin \xi _{2} \\ 
\sin \xi _{1}\sin \xi _{2} & \cos \xi _{1} & \cos \xi _{2}\sin \xi _{1}%
\end{array}%
\right) \allowbreak .
\end{equation}

Similarly, the sterile neutrino mass matrix satisfies 
\begin{equation}
M_{\nu }^{\left( 2\right) }\left( M_{\nu }^{\left( 2\right) }\right)
^{T}\simeq zv_{\chi }^{2}\left( 
\begin{array}{ccc}
a_{1}^{2}+p_{2}^{2}+p_{3}^{2} & a_{1}a_{2}+p_{1}p_{2} & a_{1}a_{3}-p_{1}p_{3}
\\ 
a_{1}a_{2}+p_{1}p_{2} & a_{2}^{2}+p_{1}^{2}+p_{3}^{2} & a_{2}a_{3}+p_{2}p_{3}
\\ 
a_{1}a_{3}-p_{1}p_{3} & a_{2}a_{3}+p_{2}p_{3} & a_{3}^{2}+p_{1}^{2}+p_{2}^{2}%
\end{array}%
\right) ,
\end{equation}%
where%
\begin{equation}
p_{j}=i\frac{v_{\rho }}{\sqrt{z}v_{\chi }}b_{j},\qquad \qquad j=1,2,3,
\end{equation}%
and $b_{j}$ are purely imaginary.

Furthermore, following the same procedure used for the light active
neutrinos, we get that the squared sterile neutrino masses are given by
\begin{eqnarray}
M_{1}^{2} &\simeq &\theta ^{2}\left[ a_{1}^{2}+a_{2}^{2}+a_{3}^{2}-\frac{%
4a_{2}^{2}\left( a_{1}^{2}+a_{3}^{2}\right) }{\left(
a_{1}^{2}+a_{2}^{2}+a_{3}^{2}\right) }\right] zv_{\chi }^{2},  \notag \\
M_{2}^{2} &\simeq &\theta ^{2}\left( a_{1}^{2}+a_{2}^{2}+a_{3}^{2}\right)
zv_{\chi }^{2},  \notag \\
M_{3}^{2} &\simeq &\left( a_{1}^{2}+a_{2}^{2}+a_{3}^{2}\right) zv_{\chi
}^{2}.
\end{eqnarray}%
where: 
\begin{equation}
\theta _{j}=\frac{p_{j}}{a_{j}}=\frac{\sqrt{2}v_{\rho }b_{j}}{\sqrt{z}%
v_{\chi }h_{\Phi j}^{\left( L\right) }}=\theta ,\qquad j=1,2,3.
\end{equation}


\end{document}